\documentclass[twocolumn,reprint,amsmath,amssymb,a4paper,aps,pra,10pt,showpacs]{revtex4}
\usepackage{amssymb}
\usepackage{amsfonts}
\usepackage{amsmath}
\usepackage{amsthm}
\usepackage{graphicx}
\usepackage{braket}
\usepackage{bm}
\usepackage{epstopdf}
\usepackage{url}
\begin{document}

\title{Suppressing the geometric dephasing of Berry phase by using modified dynamical decoupling sequences}

\author{Xiao-Ke Qin}
\author{Guang-Can Guo}
\author{Zheng-Wei Zhou}
\email[]{zwzhou@ustc.edu.cn}
\affiliation{Key Laboratory of Quantum Information, University of Science and Technology of China,\\
Chinese Academy of Sciences, Hefei, Anhui 230026, China}
\affiliation{Synergetic Innovation Center of Quantum Information and Quantum Physics, University of Science and Technology of China, Hefei, Anhui 230026, China}

\date{\today}

\begin{abstract}
Even though the traditional dynamical decoupling methods have the ability to resist dynamic dephasing caused by low frequency noise, they are not appropriate for suppressing the residual geometric dephasing, which arises from the disturbance for the geometric loop in the parameter space. This prevents the precision of quantum manipulation based geometric quantum gates from being promoted further. In this paper, we design two kinds of modified dynamical decoupling schemes to suppress the residual geometric dephasing. The further numerical simulation demonstrates the validity of our schemes.
\end{abstract}

\pacs{42.50.-p, 03.67.Lx}

\maketitle

\section{Introduction}
Berry phase\cite{berry1984} associated with the adiabatic evolution is a promising way to realize the precise control of quantum systems. By using spin echo (SE) methods\cite{jones2000geometric} one can eliminate the dynamic phase included in the total phase\cite{xiao2010} to realize the pure geometric quantum gate\cite{leek2007}. Up to now, nuclear magnetic-resonance (NMR)\cite{ota2009,du2003observation}, ultracold neutrons\cite{filipp2009}, graphene\cite{zhang2005experimental}, superconducting circuit-QED\cite{falci2000detection,falci2003quantum}, etc. have been used to study the Berry phase. The concept of geometric phase has been also generalized to Aharonov-Anandan phase\cite{aharonov1987}, non-unitary evolution\cite{carollo2003}, and mixed state\cite{tong2004}.

In the field of quantum computation, decoherence induced by the noise restricts the precision of quantum manipulation and the number of attainable quantum gates \cite{shnirman2002noise,chirolli2008,preskill1998,koh2012}. Quantum gates based on adjusting the Berry phase, ascribe to its geometric stability, are believed to have the ability to overcome certain kinds of random noise\cite{solinas2004robustness,leibfried2003experimental}. However, recent studies exhibit that the adiabatic geometric phase gates are sensitive to the noise in the control parameters \cite{blais2003effect,de2003berry}. Although dynamical decoupling methods \cite{bylander2011,cywinski2008,berger2013}, being used for adiabatic geometric phase gates, suppress the dynamic dephasing successfully, they do not work in resisting the geometric dephasing\cite{berger2015measurement,de2003berry}, which comes from the disturbance for the geometric loop in the parameter space.
How to suppress the residual geometric dephasing becomes very vital for enhancing the accuracy of the quantum gates further.

In this paper, we reinvestigate a single qubit system driven by slowly varying Hamiltonian. We simulate the classical fluctuation field along z-axis as
Ornstein-Uhlenbeck (OU) process\cite{berger2013}. By analyzing the phase difference in the adiabatic evolution, we clarify the origin leading to the residual
geometric dephasing. For instance, in the spin echo scheme, to eliminate the dynamic phase and to accumulate the Berry phase at the same time, the rotation directions of the magnetic field keep reversed before and after the flipping pulse is applied. This leads to the asymmetry of the parameters related to the low frequency noise. We design two new kinds of dynamical decoupling schemes to offset the asymmetry of noise coefficient and to suppress the residual geometric dephasing. The further numerical simulation demonstrate the validity of our schemes.

\section{Dephasing of the Berry phase}

\subsection{Berry phase}
\begin{figure}
\centering
\includegraphics[width=0.26\textwidth]{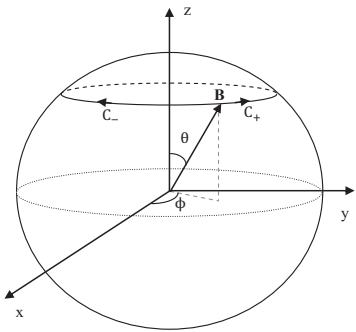}
\caption{The controllable field $\bm{B}$, which varies along the closed anti-clockwise loop $C_{+}$ or the closed clockwise loop $C_{-}$.}
\label{showspace}
\end{figure}

The qubit is driven by the Hamiltonian
\begin{equation}
H=\bm{B}(t)\cdot\bm{\sigma}/2.
\end{equation}
Here, the controllable field $\bm{B}(t)=(B_1,B_2,B_3)=B_0(t){\bf n}(t)$ and $\bm{\sigma}=(\sigma_x,\sigma_y,\sigma_z)$, where the unit direction vector ${\bf n}(t)=(\sin\theta\cos\phi,\sin\theta\sin\phi,\cos\theta)$. For simplicity, in this paper we set $\hbar=1$. By using the unitary transformation $U(t)=\exp(i\theta\sigma_y/2)\exp(i\phi\sigma_z/2)\exp(i\phi I/2)$, we obtain the Hamiltonian in the rotating frame\cite{whitney2010suppression,whitney2005},
\begin{equation}\label{rotatingpic}
\bar{H}=U(t)HU^\dag(t)+i\dot{U}(t)U^\dag(t)=\frac{1}{2}(\bar{\bm{B}}-\bar{\bm{\omega}})\cdot\bm{\sigma}-\frac{1}{2}\dot{\phi}I,
\end{equation}
where $\bar{\bm B}=(0,0,B_0)$ and $\bar{\bm{\omega}}=(-\dot{\phi}\sin\theta,\dot{\theta},\dot{\phi}\cos\theta)$.
In the ideal situation, we assume that ${\bm B}(t)$ traverses the closed anti-clockwise(clockwise) loop $C_+(C_-)$ in the time $T$,
where $B_0=const$, $\theta=const$, and $\phi$ changes from $0$ to $2 m\pi$ with the integer winding number $m$ as shown in Fig. \ref{showspace}.
Under the uniform rotation, i.e. $\dot{\phi}=2m\pi/T=\omega_{\text{rf}}$, the adiabatic condition is $|\omega_{\text{rf}}/B_0|\ll 1$\cite{de2003berry}. When $\bm{B}(t)$ does the anti-clockwise(clockwise) rotation about z-axis, $\omega_{\text{rf}}>0$ ($\omega_{\text{rf}}<0$). We expand $\Omega=|\bar{\bm{B}}-\bar{\bm{\omega}}|$ as follow
\begin{equation}\label{eigen}
\begin{split}
\Omega&=\sqrt{(B_0-\omega_{\text{rf}}\cos\theta)^2+\omega_{\text{rf}}^2\sin^2\theta}\\
      &\approx{B_0}-\omega_{\text{rf}}\cos\theta+\frac{\omega_{\text{rf}}^2\sin^2\theta}{2B_0}+o(\omega_{\text{rf}}^2/B_0).
\end{split}
\end{equation}
After the time of duration $T$, the phases of the eigenstates of the Hamiltonian in Eq. (\ref{rotatingpic}) (with the eigenvalue $\pm1$) can be calculated as:
\begin{equation}\label{gammapm}
\begin{split}
\gamma_\pm^0&=\mp\frac{1}{2}\int_0^T \Omega dt+\frac{1}{2}\int_0^T\omega_{\text{rf}} dt \\
            &\approx \mp\frac{1}{2}B_0 T+m\pi(1\pm\cos\theta)\mp\frac{m\pi\omega_{\text{rf}}}{2B_0}\sin^2\theta+\cdots.
\end{split}
\end{equation}
Here, the first term refers to the dynamic phase (proportional to T), the second one is the Berry phase (independent on T) and the third one is the non-adiabatic correction (proportional to 1/T) respectively.

Berry phase here also can be obtained in natural frame. Because the intersection angle between the vector $\bar{\bm{B}}$ and $\bar{\bm{B}}-\bar{\bm{\omega}}$ satisfies $\delta=\arctan\frac{\omega_{\text{rf}}\sin\theta}{B_0-\omega_{\text{rf}}\cos\theta}\approx 0$, in the rotating frame the eigenstates of the Hamiltonian in Eq. (\ref{rotatingpic}) can be approximated as $|{\bf n}_+\rangle=(1,0)^T$ and $|{\bf n}_-\rangle=(0,-1)^T$. Here, $|{\bf n}_+\rangle$ and $|{\bf n}_-\rangle$ are the eigenstates of $\frac{1}{2}(\bar{\bm{B}}-\bar{\bm{\omega}}_z)\cdot\bm{\sigma}$, and $\bar{\bm{\omega}}_z$ refers to the $z$ component of the vector $\bar{\bm{\omega}}$. Thus, in natural frame, the eigenstates of the operator ${\bf n}(t)\cdot\bm{\sigma}$ with the eigenvalue $s=\pm1$ can be written as:
\begin{subequations}
\begin{equation}
|\psi_{+1}({\bm{B}(t)})\rangle=U^\dag(t)|{\bf n}_+\rangle=\begin{pmatrix}
e^{-i\phi}\cos\frac{\theta}{2}\\ \sin\frac{\theta}{2}
\end{pmatrix},
\end{equation}
\begin{equation}
|\psi_{-1}({\bm{B}(t)})\rangle=U^\dag(t)|{\bf n}_-\rangle=\begin{pmatrix}
e^{-i\phi}\sin\frac{\theta}{2}\\ -\cos\frac{\theta}{2}
\end{pmatrix}.
\end{equation}
\end{subequations}
Berry phase can be represented as \cite{xiao2010,de2003berry}
\begin{equation}
\gamma^{\text{g}}_\pm=i\oint_C \langle \psi_{\pm1}({\bm{B}})|\nabla_{\bm{B}}|\psi_{\pm1}({\bm{B}})\rangle d\bm{B}=m\pi(1\pm\cos\theta).
\end{equation}

\subsection{Observation of the Berry phase}\label{obser}
To observe the Berry phase, it is necessary to induce the interference between two relevant phases. In the natural frame, assume that the initial superposition state $|\psi(0)\rangle=1/\sqrt{2}\sum_{s=\pm1}|\psi_s({\bm B}(0))\rangle$ adiabatically evolute to the final state $|\psi(T)\rangle=1/\sqrt{2} \sum_{s=\pm1}e^{i\gamma_s}|\psi_s({\bm B}(0))\rangle$. Here, the phase difference $\gamma=\gamma_{-1}-\gamma_{+1}$ is the physical observable, which contains the contributions from the dynamic phase, Berry phase and the other corrections. To observe the phase difference caused by Berry phase, ones usually take advantage of SE technique\cite{jones2000geometric,filipp2009,ota2009,leek2007,falci2000detection} to eliminate the part of the dynamic phase (here, we omit the correction term). For instance, to obtain the phase difference of Berry phase for the close anti-clockwise loop $C_+$ ($m=2$), ones first adiabatically evolves the Hamiltonian $H$ a round along anti-clockwise loop $C_+$ ($m=1$), next, a $\pi$ pulse is added to swap the eigenstates of the Hamiltonian $H$, then, an adiabatic evolution along the clockwise loop $C_-$ ($m=-1$) is applied, finally, a $\pi$ pulse is used again to swap the eigenstates of the Hamiltonian $H$. In this process, by virtue of the swapping of eigenstates, the dynamic phase is offset. Thus $\gamma_{\pm1}=\gamma_\pm^0({m=1})+\gamma_\mp^0({m=-1})=\pm2\pi\cos\theta$. The observable phase difference $\gamma=\gamma_{-1}-\gamma_{+1}=\gamma^{\text{g}}_--\gamma^{\text{g}}_+=-4\pi\cos\theta$ is pure geometric.

Based on the above protocol, we find that the final phase $\gamma_s$ is the accumulation of the phases from segmented adiabatic evolution. For the convenience of the following context we rewrite the phase $\gamma_s (s=\pm1)$ as:
\begin{equation}
\gamma_s=\sum_k \gamma_{s_k}^k(C_{l_k}^{\theta_k,s_k}).
\end{equation}
Here, the total evolution time $T$ is divided into $n$ segements: $T=\sum_{k=1}^{n}\int_{T_{k-1}}^{T_{k}} dt$ where the adjacent time intervals are separated by the swapping pulse and $s=s_k$ ($s=-s_k$) for the odd (even) $k$.  $\gamma_{s_k}^k(C_{l_k}^{\theta_k,s_k})$ is the phase integral for the eigenstate of the operator ${\bf n}(t)\cdot\bm{\sigma}$ with the eigenvalue $s_k$ in the $kth$ time interval $[T_{k-1}, T_k]$ and $C_{l_k}^{\theta_k,s_k}$ refers to the trajectory of the vector $s_k {\bf n}(t)$ in the Bloch sphere, $\theta^k$ is the intersection angle between the vector ${\bf n}(t)$ and z-axis in the $kth$ time interval.
\begin{equation}
\gamma_{s_k}^k(C_{l_k}^{\theta_k,s_k})=-\frac{s_k}{2}\int_{T_{k-1}}^{T_{k}} \Omega dt+\pi l_k,
\end{equation}
where $l_k=\omega_{\text{rf}}^k(T_k-T_{k-1})/2\pi$, $\omega_{\text{rf}}^k$ is the circular frequency in the $kth$ trajectory $C_{l_k}^{\theta_k,s_k}$ and $\omega_{\text{rf}}^k>0 (\omega_{\text{rf}}^k<0)$ refers to the anti-clockwise(clockwise) rotation about $z$ axis.

\subsection{Dephasing due to the low-frequency classical fluctuating field}
Consider the classical fluctuating field $\bm{K}(t)=(K_1,K_2,K_3)$, where $K_i, (i=1,2,3)$ is the statistical variable. The total Hamiltonian of the qubit can be written as $H_T={\bm{B}_T}(t)\cdot\bm{\sigma}/2$ with $\bm{B}_T(t)=\bm{B}(t)+\bm{K}(t)$. For simplicity, we follow the same assumption in reference \cite{leek2007,filipp2009,whitney2010suppression,whitney2005} $\bm{K}(t)=(0,0,K_3)$, which is named as $K_3$ noise and independent of the controllable field ${\bm B(t)}$. For $K_3$ noise as OU processes\cite{berger2013} with a Lorentzian spectrum of bandwidth $\Gamma_{3}$ and the noise power $\alpha_{3}$, it follows
\begin{equation}\label{corrfun}
L_{3}(\tau)=\langle K_{3}(t+\tau)K_{3}(t)\rangle=\alpha_{3}e^{-\Gamma_{3} |\tau|},
\end{equation}
which produces the power spectrum
\begin{equation}\label{spectrum}
S_{3}(\omega)={2\alpha_{3} \Gamma_{3}}/({\Gamma_{3}^2+\omega^2}).
\end{equation}
Here, we assume that the noise bandwidth $\Gamma_{3}$ is much smaller than $B_0$, which prevents the incoherent transition between two eigenstates.

Due to the existence of the classical fluctuating field, as far as the Hamiltonian in the rotating frame is concerned, it will be revised by replacing ${\bf \bar B}$ by ${\bar{\bf B}}_T=(-K_3\sin\theta,0,B_0+K_3\cos\theta)$. Thereby, we may also obtain the expansion of $\Omega_T=|\bar{\bm{B}}_T-\bar{\bm{\omega}}|$,
\begin{equation}\label{k3expansion}
\begin{split}
\Omega_T&=\left[(B_0-\omega_{\text{rf}}\cos\theta+K_3\cos\theta)^2+(\omega_{\text{rf}}-K_3)^2\sin^2\theta\right]^{\frac{1}{2}}\\&\approx\Omega+\cos\theta K_3-\frac{\omega_{\text{rf}}\sin^2\theta}{B_0}K_3+o(K_3^2/B_0),
\end{split}
\end{equation}
where the coefficients $\cos\theta$ and $-{\omega_{\text{rf}}\sin^2\theta}/{B_0}$ originate from the disturbances of the dynamic phase and the Berry phase respectively.

When $|K_3(t)|\ll B_0$ and $|\omega_{\text{rf}}|\ll B_0$, the vector $\bar{\bm{B}}_T-\bar{\bm{\omega}}$ is approximate along the same direction with the vector $\bar{\bm{B}}$.
Since there is the classical fluctuation field the phase difference $\gamma$ produces a statistical distribution, which causes the dephasing\cite{de2003berry}. In Eq. (\ref{k3expansion}), if we only consider the linear terms of $K_{3}$, the nondiagonal element $\rho_{-1,1}(T)$ in the final density matrix has the form:
\begin{equation}
\rho_{-1,1}(T)=We^{i\langle\gamma\rangle}\rho_{-1,1}(0)
\end{equation}
with the coherence function\cite{bergli2009decoherence}
\begin{equation}\label{coherence}
W=\langle e^{i \varphi}\rangle=e^{-\langle \varphi^2\rangle/2},
\end{equation}
where
\begin{equation}\label{randomp}
\varphi=-\sum_{k=1}^n\int_{T_{k-1}}^{T_{k}} s_k(\cos\theta-{\omega_{\text{rf}}^k\sin^2\theta}/{B_0}) K_{3}(t) dt
\end{equation}
with $\quad s_1=-1$. Generally, the coefficient $s_k(\cos\theta-{\omega_{\text{rf}}^k\sin^2\theta}/{B_0})$ in the adiabatic trajectory $C_{l_k}^{\theta_k,s_k}$ is unchanged. The term related to the coefficient $s_k \cos\theta$ gives rise to the dynamic dephasing (disturbance of $|\bm{B}_T|$), and the one related to the coefficient $-s_k {\omega_{\text{rf}}^k\sin^2\theta}/{B_0}$ induces the geometric dephasing\cite{de2003berry,leek2007}. We define $\chi=-\ln W=\langle \varphi^2\rangle/2$ to characterize the dephasing rate. For simplicity, we assume the Bloch vector keeps the same rate of rotation $\omega_B=|\omega_{\text{rf}}^k|$ in different trajectory $C_{l_k}^{\theta_k,s_k}$.

\begin{figure}
\centering
\includegraphics[width=0.499\textwidth]{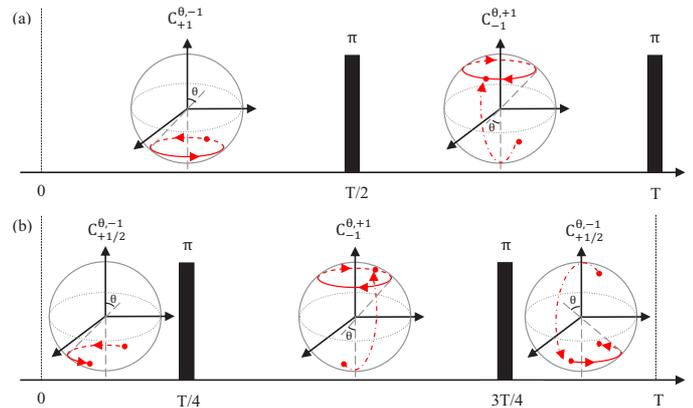}
\caption{Trajectory of the Bloch vector ${\bf b}(t)$ with the initial condition ${\bf b}(0)=-{\bf n}(0)$ as one of the eigenstate of the transient Hamiltonian $H$ in Eq. (1). The start point and end point of the trajectory segment are depicted by the red dots. Trajectories in (a) and (b) refer to $C_{+1}^{\theta,-1}C_{-1}^{\theta,+1}$ and $C_{+{1}/{2}}^{\theta,-1}C_{-1}^{\theta,+1}C_{+{1}/{2}}^{\theta,-1}$ respectively. }
\label{showspinecho}
\end{figure}

In the following context, we set the whole time of duration $T=4\pi/\omega_B$ all the time. For free induction decay (FID) (without any of a swapping
pulse), by using the correlation function in Eq. (\ref{corrfun}), we obtain the dephasing rate in the case of the low-frequency noise $\Gamma_3 T\ll 1$
\begin{equation}\label{fidsimp}
\chi\approx(\cos^2\theta-\frac{2\omega_{B}\cos\theta\sin^2\theta}{B_0}+\frac{\omega_{B}^2\sin^4\theta}{B_0^2})\frac{{\alpha_{3}}T^2}{2}.
\end{equation}
Under the adiabatic approximation, Eq. (\ref{fidsimp}) can be further reduced to:
\begin{equation}\label{fiddp}
\chi=\cos^2\theta\frac{{\alpha_{3}}T^2}{2},
\end{equation}
which is same as the dynamic dephasing of the qubit with time-independent Hamiltonian ($\theta=0$).

\section{The schemes to suppress the geometric dephasing}
\subsection{The residual geometric dephasing}\label{gdp}
Dynamical decoupling method can be usually used to suppress the low-frequency noise, which inspires its application for the Berry phase (see the Appendix). In this subsection, we will investigate the variation on dephasing rate $\chi$ by exerting traditional dynamical decoupling pulses. First, let us review SE method, which is always used to cancel the dynamic phase and suppress the dynamic dephasing\cite{filipp2009,berger2013,leek2007}. Assume that there is not any environmental noise, for the eigenstate $|\psi_{\pm 1}({\bm{B}(t)})\rangle$ of the Hamiltonian $H$, the trajectory $C_{l_1=1}^{\theta,-1}C_{l_2=-1}^{\theta,+1}$ of its Bloch vector with the initial condition ${\bf b(0)}=-{\bf n}(0)$ is exhibited in Fig. \ref{showspinecho} (a). (For simplicity, in the following all figures on the trajectory of the Bloch vector $-{\bf n}(0)$ will be acted as the starting point of the adiabatic trajectory.) As a result, for the superposition state $|\psi(t)\rangle=1/\sqrt{2}\sum_{s=\pm1}|\psi_s({\bm B}(t))\rangle$ (as shown in subsection II(B)), the dynamic phase is canceled and the Berry phase difference is accumulated, i.e. $\langle\gamma\rangle=-4\pi\cos\theta$.

When the classical fluctuating field is induced, by using Eq. (\ref{coherence}) and the correlation function in Eq. (\ref{corrfun}), we derive out
\begin{equation}\label{se}
\begin{split}
\chi=&\frac{\alpha_{3}}{\Gamma_{3}^2}\cos^2\theta(\Gamma_3 T-3+4 e^{-{\Gamma_3 T}/{2}}-e^{-\Gamma_3 T})+\\&\frac{\alpha_{3}}{\Gamma_{3}^2}\frac{\omega_{B}^2\sin^4\theta}{B_0^2}(\Gamma_{3} T-1+e^{-\Gamma_{3} T}).
\end{split}
\end{equation}
For the low-frequency noise with $\Gamma_3 T\ll 1$, we simplify the above equation as
\begin{equation}\label{sesimp}
\chi=\cos^2\theta\frac{{\alpha_{3}}\Gamma_3 T^3}{12}+\frac{\omega_{B}^2\sin^4\theta}{B_0^2}\frac{{\alpha_{3}}T^2}{2},
\end{equation}
where the first term as the dynamic dephasing is far less than Eq. (\ref{fiddp}) and the second term represents the geometric
dephasing\cite{de2003berry,berger2015measurement,leek2007}. It's obvious that SE suppresses the dynamic dephasing successfully.
However, geometric dephasing rate, which acts as one of higher order correction in Eq. (\ref{fidsimp}), can not be suppressed by SE pulses. The residual geometric dephasing becomes dominant when $\frac{\cos^2\theta}{\sin^4\theta}\frac{\beta \kappa^2}{6}\ll 1$ with $\beta=\Gamma_3 T$ (timescale of the noise) and $\kappa=B_0/\omega_B$ (adiabaticity of the evolution). The geometric dephasing does not change with $\omega_{B}$ or $T$ due to $\omega_{B} T=4\pi$. Therefore, it is not able to be suppressed by the more adiabaticity (increasing $T$), which only increases the first term.

For the more higher-order dynamical decoupling sequences, such as Carr-Purcell-Meiboom-Gill\cite{bylander2011} (CPMG) sequences,
which trajectory on Bloch vector is depicted by $C_{l_1={1}/{2}}^{\theta,-1}C_{l_2=-1}^{\theta,+1}C_{l_3={1}/{2}}^{\theta,-1}$ (see Fig. \ref{showspinecho} (b)), we may also obtain the dephasing rate as follows:
\begin{equation}\label{cpmgsimp}
\chi=\cos^2\theta\frac{{\alpha_{3}}\Gamma_3 T^3}{48}+\frac{\omega_{B}^2\sin^4\theta}{B_0^2}\frac{{\alpha_{3}}T^2}{2}.
\end{equation}
Here, although dynamic dephasing rate is further suppressed, the geometric dephasing rate is the same as that of SE.
\begin{figure}
\centering
\includegraphics[width=0.499\textwidth]{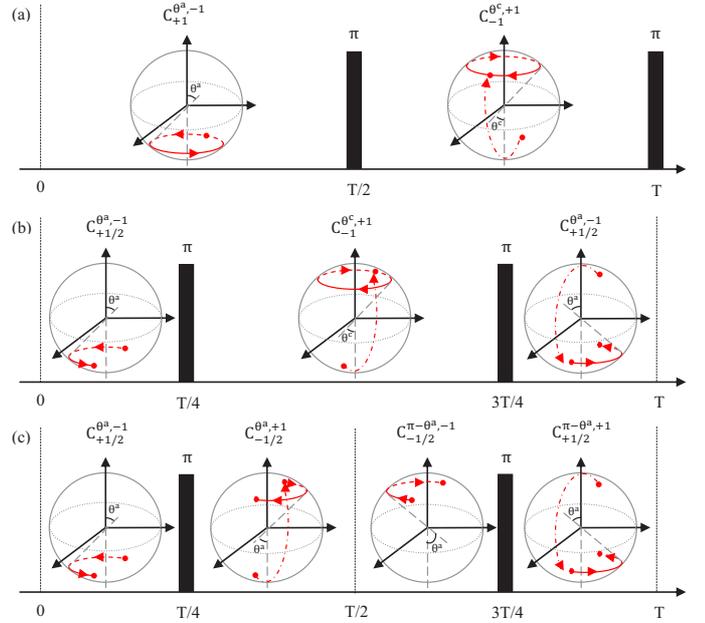}
\caption{Trajectory of the Bloch vector ${\bf b}(t)$ with the initial condition ${\bf b}(0)=-{\bf n}(0)$. The start point and end point of the trajectory segment are represented by the red dots. The trajectories $C_{+1}^{\theta^a,-1}C_{-1}^{\theta^c,+1}$, $C_{+{1}/{2}}^{\theta^a,-1}C_{-1}^{\theta^c,+1}C_{+{1}/{2}}^{\theta^a,-1}$, and $C_{+{1}/{2}}^{\theta^a,-1}C_{-1/2}^{\theta^a,+1}C_{-1/2}^{\pi-\theta^a,-1}C_{+{1}/{2}}^{\pi-\theta^a,+1}$ are depicted by (a), (b), and (c) respectively.}
\label{design}
\end{figure}

\subsection{Schemes to suppress the geometric dephasing}
To clarify why the geometric dephasing can not be further suppressed, let us look at the trajectory $C_{l_1=1}^{\theta,-1}C_{l_2=-1}^{\theta,+1}$ of the Bloch vector in SE scheme once again. Here, to eliminate the dynamical phase and to reserve the geometric phase nonzero at the same time, ones have to keep $\omega_{\text{rf}}^1=-\omega_{\text{rf}}^2$. This makes the coefficients $\cos\theta-{\omega_{\text{rf}}^1\sin^2\theta}/{B_0}\neq \cos\theta-{\omega_{\text{rf}}^2\sin^2\theta}/{B_0}$ in the integral of the Eq. (\ref{randomp}). Therefore, under the low-frequency noise approximation, although $s_1=-s_2$, the integrals in the path $C_{l_1=1}^{\theta,-1}$ and $C_{l_2=-1}^{\theta,+1}$ in the Eq. (\ref{randomp}) can not be canceled out, which causes the residual geometric dephasing. The similar analysis is also appropriate for the CPMG scheme.

To further reduce the geometric dephasing dominated by the low-frequency noise, a fundamental starting point relies on how to make the integral items in the Eq. (\ref{randomp}) offset. Here, we provide out two kinds of schemes to suppress the residual geometric dephasing as follows.

{\bf Scheme 1:} In this scheme, we vary the slant angle $\theta$ in the neighboring adiabatic path. By setting $\theta_1=\theta^a$ and $\theta_2=\theta^c$ and make the following equation hold:
\begin{equation}
    \cos\theta^a-\frac{\omega_{B}\sin^2\theta^a}{B_0}=\cos\theta^c+\frac{\omega_{B}\sin^2\theta^c}{B_0}.
\end{equation}
Here, $\omega_{B}=\omega_{\text{rf}}^1=-\omega_{\text{rf}}^2$.
Under the adiabatic condition, we derive out the approximation solution,
\begin{equation}
\cos\theta^c=\frac{\cos\theta^a-\frac{2\omega_{B}}{B_0}+\frac{\omega_{B}^2 \cos\theta^a}{B_{0}^2+\omega_{B}^2}}{1+\frac{\omega_{B}^2-2\omega_{B}B_0\cos\theta^a}{B_{0}^2+\omega_{B}^2}}.
\end{equation}
Therefore, we may modify the SE (CPMG) sequences as $C_{+1}^{\theta^a,-1}C_{-1}^{\theta^c,+1}$ ($C_{+{1}/{2}}^{\theta^a,-1}C_{-1}^{\theta^c,+1}C_{+{1}/{2}}^{\theta^a,-1}$).
The trajectories of the corresponding Bloch vector ${\bf b}(t)$ are shown in Fig \ref{design} (a) and (b) respectively.

Since there are different slant angles for two adjacent adiabatic paths, the final average Berry phase depends on two angles, i.e. $\langle\gamma\rangle=-2\pi (\cos\theta^a+\cos\theta^c)$.
By using the modified scheme, we suppress the dephasing rate as
\begin{equation}\label{secpmg}
\chi=(\cos\theta^a-\frac{\omega_{B}\sin^2\theta^a}{B_0})^2 \frac{{\alpha_{3}}\Gamma_3 T^3}{12}\begin{cases}1&
SE\\1/4& CPMG.\end{cases}
\end{equation}
For the low-frequency noise with $\Gamma_3 T\ll 1$, it decreases rapidly.

{\bf Scheme 2:} To ensure geometric phase difference $\langle\gamma\rangle=-4\pi\cos\theta^a$ we may also devise a bit complicated adiabatic paths separated by swapping pulses: $C_{+{1}/{2}}^{\theta^a,-1}C_{-1/2}^{\theta^a,+1}C_{-1/2}^{\pi-\theta^a,-1}C_{+{1}/{2}}^{\pi-\theta^a,+1}$. In these paths, the coefficients of the integral in Eq. (\ref{randomp}) are exhibited as follows:
\begin{equation}
\begin{split}
      &\cos\theta^a-\frac{\omega_{B}\sin^2\theta^a}{B_0}\rightarrow -\cos\theta^a-\frac{\omega_{B}\sin^2\theta^a}{B_0}\rightarrow\\ &-\cos\theta^a+\frac{\omega_{B}\sin^2\theta^a}{B_0}\rightarrow \cos\theta^a+\frac{\omega_{B}\sin^2\theta^a}{B_0},
\end{split}
\end{equation}
Finally, all the items are offset and the dephasing rate becomes:
\begin{equation}\label{chi1342}
\chi=\cos^2\theta^a\frac{{\alpha_{3}}\Gamma_3 T^3}{48}+\frac{\omega_{B}^2\sin^4\theta^a}{B_0^2}\frac{{\alpha_{3}}\Gamma_3 T^3}{12}.
\end{equation}
It means that the geometric dephasing is decreased by the factor $\Gamma_3 T/6$ than Eq. (\ref{sesimp}) and Eq. (\ref{cpmgsimp}).
The trajectories of the corresponding Bloch vector ${\bf b}(t)$ are shown in Fig \ref{design} (c).

\section{Numerical simulations}
In this section we numerically simulate the classical fluctuating field to check the validity for the aforementioned analytical results.
We divide the total time $T$ into $M$ intervals and keep $K_{3}(t)$ unchanged in every interval, where $M$ is set large enough so that the time step $\Delta t$ is far less than the correlation time $1/\Gamma_{3}$ of the classical fluctuation field. We generate the random noise by using the method in the references\cite{url,gillespie1996exact}
\begin{equation}\label{noise}
K_{3}(t+\Delta t)=K_{3}(t)e^{-\Gamma_{3}\Delta t}+K_{3}^{Rn}\sqrt{1-e^{-2\Gamma_{3}\Delta t}},
\end{equation}
where $K_{3}^{Rn}$ is a Gaussian distributed pseudo-random variable with the expectation $\langle K_{3}^{Rn}\rangle=0$ and the variance $\langle (K_{3}^{Rn})^2\rangle=\alpha_{3}$. The initial condition is determined by $K_{3}(0)=K_{3}^{Rn}$.

To satisfy the adiabatic approximation we take $B_0/\omega_{B}=12$, where $B_0=2\pi\nu$ and $\nu=100 \text{MHz}$. Meanwhile, considering the low-frequency noise limit $\Gamma_3/B_0\ll 1$, we set $\Delta t=(2\pi/B_0)/10$ as the minimum time scale. We also set $\Gamma_3=\beta/T$ and the variance of the classical fluctuation field $\alpha={\eta}/{\Gamma_3 T^3}$, where $\beta$ and $\eta$ are dimensionless parameters.

We start with the initial condition $|\psi(0)\rangle$ and numerically simulate the evolution of quantum state under classical fluctuating field. We repeat the calculation procedure 400 times until the average of the final density matrix ($\langle\rho(T)\rangle$) is convergent. The observable phase difference and coherence are given by $\langle\gamma\rangle=arg\langle\rho_{-1,1}(T)\rangle$ and $W=|\langle\rho_{-1,1}(T)\rangle|/|\rho_{-1,1}(0)|$ respectively.

\subsection{Results on different decoupling schemes}
\begin{figure}
\centering
\includegraphics[width=0.499\textwidth,height=0.24\textheight]{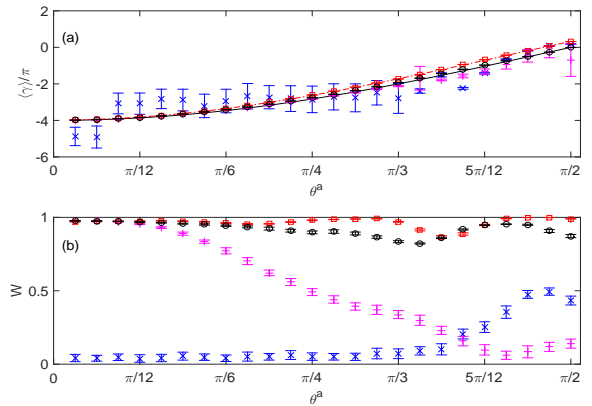}
\caption{(Color online) Numerical results under the various sequences: $C_{+2}^{\theta^a,-1}$ (blue cross), $C_{+{1}/{2}}^{\theta^a,-1}C_{-1}^{\theta^a,+1}C_{+{1}/{2}}^{\theta^a,-1}$ (magenta plus), $C_{+{1}/{2}}^{\theta^a,-1}C_{-1}^{\theta^c,+1}C_{+{1}/{2}}^{\theta^a,-1}$ (red square) and $C_{+{1}/{2}}^{\theta^a,-1}C_{-1/2}^{\theta^a,+1}C_{-1/2}^{\pi-\theta^a,-1}C_{+{1}/{2}}^{\pi-\theta^a,+1}$ (black circle). (a) The phase difference $\langle\gamma\rangle$ versus $\theta^a$, where the theoretical Berry phase difference $-4\pi\cos\theta^a$ and $-2\pi\cos\theta^a-2\pi\cos\theta^c$ are described by the solid line and dot-dashed line respectively; (b) Coherence function $W$ versus $\theta^a$. The error bars indicate the standard deviation. }
\label{k3longmemory}
\end{figure}
Under low frequency noise, we simulate the evolutions of quantum system for different decoupling schemes characterized by the trajectories $C_{+2}^{\theta^a,-1}$, $C_{+{1}/{2}}^{\theta^a,-1}C_{-1}^{\theta^a,+1}C_{+{1}/{2}}^{\theta^a,-1}$, $C_{+{1}/{2}}^{\theta^a,-1}C_{-1}^{\theta^c,+1}C_{+{1}/{2}}^{\theta^a,-1}$ and $C_{+{1}/{2}}^{\theta^a,-1}C_{-1/2}^{\theta^a,+1}C_{-1/2}^{\pi-\theta^a,-1}C_{+{1}/{2}}^{\pi-\theta^a,+1}$ respectively. Here, we set $\beta=0.001$ and $\eta=400\beta$.
The phase difference $\langle\gamma\rangle$ for different decoupling schemes are displayed in Fig. \ref{k3longmemory} (a), where the
theoretical Berry phase difference $-4\pi\cos\theta^a$ and $-2\pi\cos\theta^a-2\pi\cos\theta^c$ are described by the solid line and
dot-dashed line respectively. Fig. \ref{k3longmemory} (b) exhibits different coherence functions $W$ versus $\theta$. For CPMG scheme
characterized by $C_{+{1}/{2}}^{\theta^a,-1}C_{-1}^{\theta^a,+1}C_{+{1}/{2}}^{\theta^a,-1}$, the coherence function $W$ decays as $\theta^a$ grows.
It is because the residual geometric dephasing becomes dominant when $|\sin\theta^a|$ increases. While, for the modified decoupling schemes,
both $C_{+{1}/{2}}^{\theta^a,-1}C_{-1}^{\theta^c,+1}C_{+{1}/{2}}^{\theta^a,-1}$ and
$C_{+{1}/{2}}^{\theta^a,-1}C_{-1/2}^{\theta^a,+1}C_{-1/2}^{\pi-\theta^a,-1}C_{+{1}/{2}}^{\pi-\theta^a,+1}$ suppress the geometric dephasing successfully,
which improve the coherence and recover the Berry phases at the whole angle range.

It is slightly strange that the coherence for FID scheme is better than that for CPMG in large $\theta$. The reason is that CPMG scheme makes the second term in in Eq. (\ref{fidsimp}) disappear which originally resists the residual geometric dephasing.

\subsection{The correlation time}
\begin{figure}
\centering
\includegraphics[width=0.499\textwidth,height=0.24\textheight]{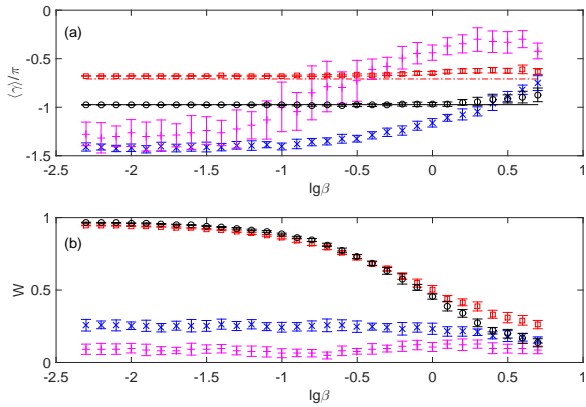}
\caption{(Color online) The Berry phase difference $\langle\gamma\rangle$ (a) and coherence function $W$ (b) versus $\beta=\Gamma_3 T$ on the condition of $\theta^a=5\pi/12$, where all the labels are same as Fig. \ref{k3longmemory}. }
\label{k3memorydepen}
\end{figure}
We also investigate the suppressing performance of various decoupling schemes versus the correlation time of the low-frequency fluctuating field. Here, we fix the angle $\theta^a=5\pi/12$. By keeping the total evolution time $T$ unchanged we may use $\beta=\Gamma_3 T$ to adjust the correlation time of $K_{3}$, where $\beta\in [0.005,5]$.

In Fig. \ref{k3memorydepen}, we depict the phase difference and the coherence function $W$ versus $\lg\beta$. 
The modified decoupling schemes $C_{+{1}/{2}}^{\theta^a,-1}C_{-1}^{\theta^c,+1}C_{+{1}/{2}}^{\theta^a,-1}$ and $C_{+{1}/{2}}^{\theta^a,-1}C_{-1/2}^{\theta^a,+1}C_{-1/2}^{\pi-\theta^a,-1}C_{+{1}/{2}}^{\pi-\theta^a,+1}$ suppress the geometric dephasing until $\beta=\Gamma_3 T\sim 1$. If the correlation time is less than the evolution time $\beta=\Gamma_3 T>1$, any dynamical decoupling schemes fail.

\section{Discussion and Summary}
Generally speaking, dephasing is not the only decoherence source in our system when the noise bandwidth $\Gamma_3$ becomes large, i.e. $\Gamma_3\sim B_0$. It induces non-adiabatic transition, which gives the depolarization factor $\exp{(-\lambda)}$ and the coherence function $W=\exp{(-\lambda/2-\chi)}$\cite{ithier2005decoherence}, where $\lambda=\alpha_3 T\sin^2\theta \Gamma_3/(\Gamma_3^2+B_0^2)$. In this situation, this kind of strategy of adiabatic driving accompanied by dynamical decoupling sequences does not work.

The foregoing schemes do not involve the case of the transverse noise. Recently, the artificial simulated transverse noise with $K_1=K_r\cos\phi$ and $K_2=K_r \sin\phi$ has also been studied\cite{berger2015measurement}, where $K_r$ is the magnetic field fluctuation in radial direction and it originates from the driving amplitude noise of qubit. Similarly, in the rotating frame, we obtain ${\bar{\bf B}}_T=(K_r\cos\theta,0,B_0+K_r\sin\theta)$ and
 \begin{equation}\label{krexpansion}
\begin{split}
\Omega_T&=\left[(B_0-\omega_{\text{rf}}\cos\theta+K_r\sin\theta)^2+(\omega_{\text{rf}}\sin\theta+K_r\cos\theta)^2\right]^{\frac{1}{2}}\\&\approx\Omega+\sin\theta K_r-\frac{\omega_{\text{rf}}\cos\theta\sin\theta}{B_0}K_r+o(K_r^2/B_0),
\end{split}
\end{equation}
Following the similar procedure discussed in scheme 1, we can also suppress the residual geometric dephasing except $\theta^a$ and $\theta^c$ satisfy:
\begin{equation}
    \sin\theta^a(1-\frac{\omega_{B}\cos\theta^a}{B_0})=\sin\theta^c(1+\frac{\omega_{B}\cos\theta^c}{B_0}).
\end{equation}
However, in this situation, the strategy in scheme 2 does not work because $\omega_{\text{rf}}\cos\theta$ and $\omega_{\text{rf}}\cos\theta\sin\theta/B_0$ have the same sign. In addition, how to design dynamical decoupling sequences for overcoming residual geometric dephasing induced by both longitudinal noise and transverse noise simultaneously remains open.

In summary, we have studied how to suppress the geometric dephasing of Berry phase induced by the classical low-frequency fluctuation field $K_{3}(t)$. By taking advantage of the expansion of total Hamiltonian in the rotating frame, we analyze the origin of the residual geometric dephasing. In essence, it stems from the time-varying intersection angle between the noise and the effective Hamiltonian with related to the rotation direction, which can not be suppressed further by using the traditional dynamical decoupling strategies. Furthermore, we present two kinds of modified dynamical decoupling schemes to balance the noises in different time interval and suppress the residual geometric dephasing. Numerical results verify the validity of our schemes. These schemes will help to generate more high-precision quantum manipulations and to restrain decoherence induced by the low-frequency noise.

\begin{acknowledgments}
This work was funded by the National Key Research and Development Program (Grant No.2016YFA0301700), the National Natural Science Foundation of China (Grant No. 11574294, 61490711), and the Strategic Priority Research Program of the Chinese Academy of Sciences (Grant No. XDB01030200).
\end{acknowledgments}

\appendix*
\section{Dynamical decoupling}\label{dd}

Dynamical decoupling is used to suppress the low-frequency noise through flipping the coefficients of random noise rapidly. Finally, the random variable in the phase difference is
\begin{equation}\label{randomphase}
\varphi=\int_0^{T} h(t,T)K_3(t) dt
\end{equation}
with the function $h(t,T)=\pm 1$ determined by
\begin{equation}
h(t,T)=\sum_{k=0}^{m}(-1)^k\Theta(t_{k+1}-t)\Theta(t-t_{k}),
\end{equation}
where $\Theta(t)$ is the Heaviside step function, $t_0=0$, $t_{m+1}=T$, and $m$ represents the total times of switching between $1$ and $-1$. The time $t_k$ depends on the dynamical decoupling sequences in detail.

\begin{table}
\caption{$h(t,T)$ and the corresponding filter functions of several dynamical decoupling sequences.}
\label{tab1}
\centering
\begin{ruledtabular}
\begin{tabular}{c|c|c}
$h(t,T)$&Sequence&F(z)\\
\hline
++++&FID& $2\sin^2\frac{z}{2}$\\
\hline
+ + - -&SE& $8\sin^4\frac{z}{4}$\\
\hline
+ - - +&CPMG(n=2)& $8\sin^4\frac{z}{4n}\sin^2\frac{z}{2}/\cos^2\frac{z}{2n}$\\
\end{tabular}
\end{ruledtabular}
\end{table}

Using Eq. (\ref{coherence}), the dephasing is represented by\cite{cywinski2008}
\begin{equation}\label{chit}
\chi=-\ln W=\int_0^\infty\frac{d\omega}{\pi}S_3(\omega)\frac{F(\omega T)}{\omega^2}.
\end{equation}
In Table \ref{tab1}, we summarize $h(t,T)$ and the corresponding filter functions $F$ of several dynamical decoupling sequences, where $\pm$ refers to the value of the function $h(t,T)$ in every time of duration $T/4$. For example, $++++$ refers to $h(t,T)=1$ in the whole time of duration $T$. Using the spectrum in Eq. (\ref{spectrum}), we derive out
\begin{widetext}
\begin{equation}\label{effect}
\chi(T)=\frac{\alpha}{\Gamma_3^2}
\begin{cases}
\Gamma_3 T-1+e^{-\Gamma_3 T}& \text{FID}\\
\Gamma_3 T-3+4 e^{-{\Gamma_3 T}/{2}}-e^{-\Gamma_3 T}& \text{SE}\\
\Gamma_3 T-5+4 e^{-{\Gamma_3 T}/{4}}+4 e^{-{\Gamma_3 T}/{2}}-4 e^{-{3\Gamma_3 T}/{4}}+e^{-\Gamma_3 T}& \text{CPMG (n=2)}\\
\end{cases}
\end{equation}
\end{widetext}

\begin{figure}
\centering
\includegraphics[width=0.38\textwidth,height=0.13\textheight]{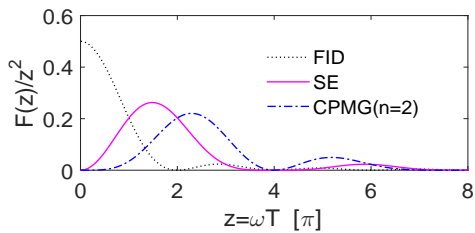}
\caption{(Color online) $F(z)/z^2$ vs $z$ for various dynamical decoupling sequences. The frequency with the maximum $F(z)/z^2$ increases as $h(t,T)$ switches more times.}
\label{filterfig}
\end{figure}

Now we analyze the above equation in various limiting cases. For the low-frequency noise, we obtain $\Gamma_3 T\ll 1$. Then Eq. (\ref{effect}) is rewritten as
\begin{equation}\label{effectbeta}
\chi(T)\approx\frac{\alpha T^2}{2}\begin{cases}
1 & \text{FID}\\
{\Gamma_3 T}/{6} & \text{SE}\\
{\Gamma_3 T}/{24} & \text{CPMG(n=2)}
\end{cases}
\end{equation}
where the dephasing is suppressed by dynamical decoupling sequences. We plot $F(z)/z^2$ vs z in Fig. \ref{filterfig}, where the low-frequency contribution of noise can be filtered out through some sequences. The frequency with the maximum $F(z)/z^2$ increases as $h(t,T)$ switches more times. For the high-frequency noise with $\Gamma_3 T\gg 1$, $\chi(T)\approx{\alpha T}/{\Gamma_3}$ for any sequences, which means that none of them is valid for suppressing the high-frequency noise.


\begin{thebibliography}{31}
\expandafter\ifx\csname natexlab\endcsname\relax\def\natexlab#1{#1}\fi
\expandafter\ifx\csname bibnamefont\endcsname\relax
  \def\bibnamefont#1{#1}\fi
\expandafter\ifx\csname bibfnamefont\endcsname\relax
  \def\bibfnamefont#1{#1}\fi
\expandafter\ifx\csname citenamefont\endcsname\relax
  \def\citenamefont#1{#1}\fi
\expandafter\ifx\csname url\endcsname\relax
  \def\url#1{\texttt{#1}}\fi
\expandafter\ifx\csname urlprefix\endcsname\relax\def\urlprefix{URL }\fi
\providecommand{\bibinfo}[2]{#2}
\providecommand{\eprint}[2][]{\url{#2}}

\bibitem[{\citenamefont{Berry}(1984)}]{berry1984}
\bibinfo{author}{\bibfnamefont{M.}~\bibnamefont{Berry}}, \bibinfo{journal}{P.
  Roy. Soc. A-Math. Phy.} \textbf{\bibinfo{volume}{392}}, \bibinfo{pages}{45}
  (\bibinfo{year}{1984}).

\bibitem[{\citenamefont{Jones et~al.}(2000)\citenamefont{Jones, Vedral, Ekert,
  and Castagnoli}}]{jones2000geometric}
\bibinfo{author}{\bibfnamefont{J.~A.} \bibnamefont{Jones}},
  \bibinfo{author}{\bibfnamefont{V.}~\bibnamefont{Vedral}},
  \bibinfo{author}{\bibfnamefont{A.}~\bibnamefont{Ekert}}, \bibnamefont{and}
  \bibinfo{author}{\bibfnamefont{G.}~\bibnamefont{Castagnoli}},
  \bibinfo{journal}{Nature} \textbf{\bibinfo{volume}{403}},
  \bibinfo{pages}{869} (\bibinfo{year}{2000}).


\bibitem[{\citenamefont{Xiao et~al.}(2010)\citenamefont{Xiao, Chang, and
  Niu}}]{xiao2010}
\bibinfo{author}{\bibfnamefont{D.}~\bibnamefont{Xiao}},
  \bibinfo{author}{\bibfnamefont{M.-C.} \bibnamefont{Chang}}, \bibnamefont{and}
  \bibinfo{author}{\bibfnamefont{Q.}~\bibnamefont{Niu}}, \bibinfo{journal}{Rev.
  Mod. Phys.} \textbf{\bibinfo{volume}{82}}, \bibinfo{pages}{1959}
  (\bibinfo{year}{2010}).

\bibitem[{\citenamefont{Leek et~al.}(2007)\citenamefont{Leek, Fink, Blais,
  Bianchetti, G{\"o}ppl, Gambetta, Schuster, Frunzio, Schoelkopf, and
  Wallraff}}]{leek2007}
\bibinfo{author}{\bibfnamefont{P.}~\bibnamefont{Leek}},
  \bibinfo{author}{\bibfnamefont{J.}~\bibnamefont{Fink}},
  \bibinfo{author}{\bibfnamefont{A.}~\bibnamefont{Blais}},
  \bibinfo{author}{\bibfnamefont{R.}~\bibnamefont{Bianchetti}},
  \bibinfo{author}{\bibfnamefont{M.}~\bibnamefont{G{\"o}ppl}},
  \bibinfo{author}{\bibfnamefont{J.}~\bibnamefont{Gambetta}},
  \bibinfo{author}{\bibfnamefont{D.}~\bibnamefont{Schuster}},
  \bibinfo{author}{\bibfnamefont{L.}~\bibnamefont{Frunzio}},
  \bibinfo{author}{\bibfnamefont{R.}~\bibnamefont{Schoelkopf}},
  \bibnamefont{and} \bibinfo{author}{\bibfnamefont{A.}~\bibnamefont{Wallraff}},
  \bibinfo{journal}{Science} \textbf{\bibinfo{volume}{318}},
  \bibinfo{pages}{1889} (\bibinfo{year}{2007}).

\bibitem[{\citenamefont{Ota et~al.}(2009)\citenamefont{Ota, Goto, Kondo, and
  Nakahara}}]{ota2009}
\bibinfo{author}{\bibfnamefont{Y.}~\bibnamefont{Ota}},
  \bibinfo{author}{\bibfnamefont{Y.}~\bibnamefont{Goto}},
  \bibinfo{author}{\bibfnamefont{Y.}~\bibnamefont{Kondo}}, \bibnamefont{and}
  \bibinfo{author}{\bibfnamefont{M.}~\bibnamefont{Nakahara}},
  \bibinfo{journal}{Phys. Rev. A} \textbf{\bibinfo{volume}{80}},
  \bibinfo{pages}{052311} (\bibinfo{year}{2009}).

\bibitem[{\citenamefont{Du et~al.}(2003)\citenamefont{Du, Zou, Shi, Kwek, Pan,
  Oh, Ekert, Oi, and Ericsson}}]{du2003observation}
\bibinfo{author}{\bibfnamefont{J.}~\bibnamefont{Du}},
  \bibinfo{author}{\bibfnamefont{P.}~\bibnamefont{Zou}},
  \bibinfo{author}{\bibfnamefont{M.}~\bibnamefont{Shi}},
  \bibinfo{author}{\bibfnamefont{L.~C.} \bibnamefont{Kwek}},
  \bibinfo{author}{\bibfnamefont{J.-W.} \bibnamefont{Pan}},
  \bibinfo{author}{\bibfnamefont{C.~H.} \bibnamefont{Oh}},
  \bibinfo{author}{\bibfnamefont{A.}~\bibnamefont{Ekert}},
  \bibinfo{author}{\bibfnamefont{D.~K.} \bibnamefont{Oi}}, \bibnamefont{and}
  \bibinfo{author}{\bibfnamefont{M.}~\bibnamefont{Ericsson}},
  \bibinfo{journal}{Phys. Rev. Lett.} \textbf{\bibinfo{volume}{91}},
  \bibinfo{pages}{100403} (\bibinfo{year}{2003}).

\bibitem[{\citenamefont{Filipp et~al.}(2009)\citenamefont{Filipp, Klepp,
  Hasegawa, Plonka-Spehr, Schmidt, Geltenbort, and Rauch}}]{filipp2009}
\bibinfo{author}{\bibfnamefont{S.}~\bibnamefont{Filipp}},
  \bibinfo{author}{\bibfnamefont{J.}~\bibnamefont{Klepp}},
  \bibinfo{author}{\bibfnamefont{Y.}~\bibnamefont{Hasegawa}},
  \bibinfo{author}{\bibfnamefont{C.}~\bibnamefont{Plonka-Spehr}},
  \bibinfo{author}{\bibfnamefont{U.}~\bibnamefont{Schmidt}},
  \bibinfo{author}{\bibfnamefont{P.}~\bibnamefont{Geltenbort}},
  \bibnamefont{and} \bibinfo{author}{\bibfnamefont{H.}~\bibnamefont{Rauch}},
  \bibinfo{journal}{Phys. Rev. Lett.} \textbf{\bibinfo{volume}{102}},
  \bibinfo{pages}{030404} (\bibinfo{year}{2009}).

\bibitem[{\citenamefont{Zhang et~al.}(2005)\citenamefont{Zhang, Tan, Stormer,
  and Kim}}]{zhang2005experimental}
\bibinfo{author}{\bibfnamefont{Y.}~\bibnamefont{Zhang}},
  \bibinfo{author}{\bibfnamefont{Y.-W.} \bibnamefont{Tan}},
  \bibinfo{author}{\bibfnamefont{H.~L.} \bibnamefont{Stormer}},
  \bibnamefont{and} \bibinfo{author}{\bibfnamefont{P.}~\bibnamefont{Kim}},
  \bibinfo{journal}{Nature} \textbf{\bibinfo{volume}{438}},
  \bibinfo{pages}{201} (\bibinfo{year}{2005}).

\bibitem[{\citenamefont{Falci et~al.}(2000)\citenamefont{Falci, Fazio, Palma,
  Siewert, and Vedral}}]{falci2000detection}
\bibinfo{author}{\bibfnamefont{G.}~\bibnamefont{Falci}},
  \bibinfo{author}{\bibfnamefont{R.}~\bibnamefont{Fazio}},
  \bibinfo{author}{\bibfnamefont{G.~M.} \bibnamefont{Palma}},
  \bibinfo{author}{\bibfnamefont{J.}~\bibnamefont{Siewert}}, \bibnamefont{and}
  \bibinfo{author}{\bibfnamefont{V.}~\bibnamefont{Vedral}},
  \bibinfo{journal}{Nature} \textbf{\bibinfo{volume}{407}},
  \bibinfo{pages}{355} (\bibinfo{year}{2000}).

\bibitem[{\citenamefont{Falci et~al.}(2003)\citenamefont{Falci, Fazio, and
  Palma}}]{falci2003quantum}
\bibinfo{author}{\bibfnamefont{G.}~\bibnamefont{Falci}},
  \bibinfo{author}{\bibfnamefont{R.}~\bibnamefont{Fazio}}, \bibnamefont{and}
  \bibinfo{author}{\bibfnamefont{G.~M.} \bibnamefont{Palma}},
  \bibinfo{journal}{Fortschr. Phys.} \textbf{\bibinfo{volume}{51}},
  \bibinfo{pages}{442} (\bibinfo{year}{2003}).

\bibitem[{\citenamefont{Aharonov and Anandan}(1987)}]{aharonov1987}
\bibinfo{author}{\bibfnamefont{Y.}~\bibnamefont{Aharonov}} \bibnamefont{and}
  \bibinfo{author}{\bibfnamefont{J.}~\bibnamefont{Anandan}},
  \bibinfo{journal}{Phys. Rev. Lett.} \textbf{\bibinfo{volume}{58}},
  \bibinfo{pages}{1593} (\bibinfo{year}{1987}).

\bibitem[{\citenamefont{Carollo et~al.}(2003)\citenamefont{Carollo,
  Fuentes-Guridi, Santos, and Vedral}}]{carollo2003}
\bibinfo{author}{\bibfnamefont{A.}~\bibnamefont{Carollo}},
  \bibinfo{author}{\bibfnamefont{I.}~\bibnamefont{Fuentes-Guridi}},
  \bibinfo{author}{\bibfnamefont{M.~F.} \bibnamefont{Santos}},
  \bibnamefont{and} \bibinfo{author}{\bibfnamefont{V.}~\bibnamefont{Vedral}},
  \bibinfo{journal}{Phys. Rev. Lett.} \textbf{\bibinfo{volume}{90}},
  \bibinfo{pages}{160402} (\bibinfo{year}{2003}).

\bibitem[{\citenamefont{Tong et~al.}(2004)\citenamefont{Tong, Sj{\"o}qvist,
  Kwek, and Oh}}]{tong2004}
\bibinfo{author}{\bibfnamefont{D.}~\bibnamefont{Tong}},
  \bibinfo{author}{\bibfnamefont{E.}~\bibnamefont{Sj{\"o}qvist}},
  \bibinfo{author}{\bibfnamefont{L.~C.} \bibnamefont{Kwek}}, \bibnamefont{and}
  \bibinfo{author}{\bibfnamefont{C.}~\bibnamefont{Oh}}, \bibinfo{journal}{Phys.
  Rev. Lett.} \textbf{\bibinfo{volume}{93}}, \bibinfo{pages}{080405}
  (\bibinfo{year}{2004}).

\bibitem[{\citenamefont{Shnirman et~al.}(2002)\citenamefont{Shnirman, Makhlin,
  and Sch{\"o}n}}]{shnirman2002noise}
\bibinfo{author}{\bibfnamefont{A.}~\bibnamefont{Shnirman}},
  \bibinfo{author}{\bibfnamefont{Y.}~\bibnamefont{Makhlin}}, \bibnamefont{and}
  \bibinfo{author}{\bibfnamefont{G.}~\bibnamefont{Sch{\"o}n}},
  \bibinfo{journal}{Phys. Scripta} \textbf{\bibinfo{volume}{2002}},
  \bibinfo{pages}{147} (\bibinfo{year}{2002}).

\bibitem[{\citenamefont{Chirolli and Burkard}(2008)}]{chirolli2008}
\bibinfo{author}{\bibfnamefont{L.}~\bibnamefont{Chirolli}} \bibnamefont{and}
  \bibinfo{author}{\bibfnamefont{G.}~\bibnamefont{Burkard}},
  \bibinfo{journal}{Adv. Phys.} \textbf{\bibinfo{volume}{57}},
  \bibinfo{pages}{225} (\bibinfo{year}{2008}).

\bibitem[{\citenamefont{Preskill}(1998)}]{preskill1998}
\bibinfo{author}{\bibfnamefont{J.}~\bibnamefont{Preskill}},
  \bibinfo{journal}{P. Roy. Soc. A-Math. Phy.} \textbf{\bibinfo{volume}{454}},
  \bibinfo{pages}{469} (\bibinfo{year}{1998}).

\bibitem[{\citenamefont{Koh et~al.}(2012)\citenamefont{Koh, Gamble, Friesen,
  Eriksson, and Coppersmith}}]{koh2012}
\bibinfo{author}{\bibfnamefont{T.~S.} \bibnamefont{Koh}},
  \bibinfo{author}{\bibfnamefont{J.~K.} \bibnamefont{Gamble}},
  \bibinfo{author}{\bibfnamefont{M.}~\bibnamefont{Friesen}},
  \bibinfo{author}{\bibfnamefont{M.}~\bibnamefont{Eriksson}}, \bibnamefont{and}
  \bibinfo{author}{\bibfnamefont{S.}~\bibnamefont{Coppersmith}},
  \bibinfo{journal}{Phys. Rev. Lett.} \textbf{\bibinfo{volume}{109}},
  \bibinfo{pages}{250503} (\bibinfo{year}{2012}).

\bibitem[{\citenamefont{Solinas et~al.}(2004)\citenamefont{Solinas, Zanardi,
  and Zangh{\`\i}}}]{solinas2004robustness}
\bibinfo{author}{\bibfnamefont{P.}~\bibnamefont{Solinas}},
  \bibinfo{author}{\bibfnamefont{P.}~\bibnamefont{Zanardi}}, \bibnamefont{and}
  \bibinfo{author}{\bibfnamefont{N.}~\bibnamefont{Zangh{\`\i}}},
  \bibinfo{journal}{Phys. Rev. A} \textbf{\bibinfo{volume}{70}},
  \bibinfo{pages}{042316} (\bibinfo{year}{2004}).

\bibitem[{\citenamefont{Leibfried et~al.}(2003)\citenamefont{Leibfried,
  DeMarco, Meyer, Lucas, Barrett, Britton, Itano, Jelenkovi{\'c}, Langer, and
  Wineland}}]{leibfried2003experimental}
\bibinfo{author}{\bibfnamefont{D.}~\bibnamefont{Leibfried}},
  \bibinfo{author}{\bibfnamefont{B.}~\bibnamefont{DeMarco}},
  \bibinfo{author}{\bibfnamefont{V.}~\bibnamefont{Meyer}},
  \bibinfo{author}{\bibfnamefont{D.}~\bibnamefont{Lucas}},
  \bibinfo{author}{\bibfnamefont{M.}~\bibnamefont{Barrett}},
  \bibinfo{author}{\bibfnamefont{J.}~\bibnamefont{Britton}},
  \bibinfo{author}{\bibfnamefont{W.}~\bibnamefont{Itano}},
  \bibinfo{author}{\bibfnamefont{B.}~\bibnamefont{Jelenkovi{\'c}}},
  \bibinfo{author}{\bibfnamefont{C.}~\bibnamefont{Langer}}, \bibnamefont{and}
  \bibinfo{author}{\bibfnamefont{D.~J.} \bibnamefont{Wineland}},
  \bibinfo{journal}{Nature} \textbf{\bibinfo{volume}{422}},
  \bibinfo{pages}{412} (\bibinfo{year}{2003}).

\bibitem[{\citenamefont{Blais and Tremblay}(2003)}]{blais2003effect}
\bibinfo{author}{\bibfnamefont{A.}~\bibnamefont{Blais}} \bibnamefont{and}
  \bibinfo{author}{\bibfnamefont{A.-M.} \bibnamefont{Tremblay}},
  \bibinfo{journal}{Phys. Rev. A} \textbf{\bibinfo{volume}{67}},
  \bibinfo{pages}{012308} (\bibinfo{year}{2003}).

\bibitem[{\citenamefont{De~Chiara and Palma}(2003)}]{de2003berry}
\bibinfo{author}{\bibfnamefont{G.}~\bibnamefont{De~Chiara}} \bibnamefont{and}
  \bibinfo{author}{\bibfnamefont{G.~M.} \bibnamefont{Palma}},
  \bibinfo{journal}{Phys. Rev. Lett.} \textbf{\bibinfo{volume}{91}},
  \bibinfo{pages}{090404} (\bibinfo{year}{2003}).

\bibitem[{\citenamefont{Bylander et~al.}(2011)\citenamefont{Bylander,
  Gustavsson, Yan, Yoshihara, Harrabi, Fitch, Cory, Nakamura, Tsai, and
  Oliver}}]{bylander2011}
\bibinfo{author}{\bibfnamefont{J.}~\bibnamefont{Bylander}},
  \bibinfo{author}{\bibfnamefont{S.}~\bibnamefont{Gustavsson}},
  \bibinfo{author}{\bibfnamefont{F.}~\bibnamefont{Yan}},
  \bibinfo{author}{\bibfnamefont{F.}~\bibnamefont{Yoshihara}},
  \bibinfo{author}{\bibfnamefont{K.}~\bibnamefont{Harrabi}},
  \bibinfo{author}{\bibfnamefont{G.}~\bibnamefont{Fitch}},
  \bibinfo{author}{\bibfnamefont{D.~G.} \bibnamefont{Cory}},
  \bibinfo{author}{\bibfnamefont{Y.}~\bibnamefont{Nakamura}},
  \bibinfo{author}{\bibfnamefont{J.-S.} \bibnamefont{Tsai}}, \bibnamefont{and}
  \bibinfo{author}{\bibfnamefont{W.~D.} \bibnamefont{Oliver}},
  \bibinfo{journal}{Nat. Phys.} \textbf{\bibinfo{volume}{7}},
  \bibinfo{pages}{565} (\bibinfo{year}{2011}).

\bibitem[{\citenamefont{Cywi{\'n}ski et~al.}(2008)\citenamefont{Cywi{\'n}ski,
  Lutchyn, Nave, and Sarma}}]{cywinski2008}
\bibinfo{author}{\bibfnamefont{{\L}.}~\bibnamefont{Cywi{\'n}ski}},
  \bibinfo{author}{\bibfnamefont{R.~M.} \bibnamefont{Lutchyn}},
  \bibinfo{author}{\bibfnamefont{C.~P.} \bibnamefont{Nave}}, \bibnamefont{and}
  \bibinfo{author}{\bibfnamefont{S.~D.} \bibnamefont{Sarma}},
  \bibinfo{journal}{Phys. Rev. B} \textbf{\bibinfo{volume}{77}},
  \bibinfo{pages}{174509} (\bibinfo{year}{2008}).

\bibitem[{\citenamefont{Berger et~al.}(2013)\citenamefont{Berger, Pechal,
  Abdumalikov~Jr, Eichler, Steffen, Fedorov, Wallraff, and
  Filipp}}]{berger2013}
\bibinfo{author}{\bibfnamefont{S.}~\bibnamefont{Berger}},
  \bibinfo{author}{\bibfnamefont{M.}~\bibnamefont{Pechal}},
  \bibinfo{author}{\bibfnamefont{A.}~\bibnamefont{Abdumalikov~Jr}},
  \bibinfo{author}{\bibfnamefont{C.}~\bibnamefont{Eichler}},
  \bibinfo{author}{\bibfnamefont{L.}~\bibnamefont{Steffen}},
  \bibinfo{author}{\bibfnamefont{A.}~\bibnamefont{Fedorov}},
  \bibinfo{author}{\bibfnamefont{A.}~\bibnamefont{Wallraff}}, \bibnamefont{and}
  \bibinfo{author}{\bibfnamefont{S.}~\bibnamefont{Filipp}},
  \bibinfo{journal}{Phys. Rev. A} \textbf{\bibinfo{volume}{87}},
  \bibinfo{pages}{060303} (\bibinfo{year}{2013}).

\bibitem[{\citenamefont{Berger et~al.}(2015)\citenamefont{Berger, Pechal,
  Kurpiers, Abdumalikov, Eichler, Mlynek, Shnirman, Gefen, Wallraff, and
  Filipp}}]{berger2015measurement}
\bibinfo{author}{\bibfnamefont{S.}~\bibnamefont{Berger}},
  \bibinfo{author}{\bibfnamefont{M.}~\bibnamefont{Pechal}},
  \bibinfo{author}{\bibfnamefont{P.}~\bibnamefont{Kurpiers}},
  \bibinfo{author}{\bibfnamefont{A.}~\bibnamefont{Abdumalikov}},
  \bibinfo{author}{\bibfnamefont{C.}~\bibnamefont{Eichler}},
  \bibinfo{author}{\bibfnamefont{J.}~\bibnamefont{Mlynek}},
  \bibinfo{author}{\bibfnamefont{A.}~\bibnamefont{Shnirman}},
  \bibinfo{author}{\bibfnamefont{Y.}~\bibnamefont{Gefen}},
  \bibinfo{author}{\bibfnamefont{A.}~\bibnamefont{Wallraff}}, \bibnamefont{and}
  \bibinfo{author}{\bibfnamefont{S.}~\bibnamefont{Filipp}},
  \bibinfo{journal}{Nat. Commun.} \textbf{\bibinfo{volume}{6}},
  \bibinfo{pages}{8757} (\bibinfo{year}{2015}).

\bibitem[{\citenamefont{Whitney}(2010)}]{whitney2010suppression}
\bibinfo{author}{\bibfnamefont{R.~S.} \bibnamefont{Whitney}},
  \bibinfo{journal}{Phys. Rev. A} \textbf{\bibinfo{volume}{81}},
  \bibinfo{pages}{032108} (\bibinfo{year}{2010}).

\bibitem[{\citenamefont{Whitney et~al.}(2005)\citenamefont{Whitney, Makhlin,
  Shnirman, and Gefen}}]{whitney2005}
\bibinfo{author}{\bibfnamefont{R.~S.} \bibnamefont{Whitney}},
  \bibinfo{author}{\bibfnamefont{Y.}~\bibnamefont{Makhlin}},
  \bibinfo{author}{\bibfnamefont{A.}~\bibnamefont{Shnirman}}, \bibnamefont{and}
  \bibinfo{author}{\bibfnamefont{Y.}~\bibnamefont{Gefen}},
  \bibinfo{journal}{Phys. Rev. Lett.} \textbf{\bibinfo{volume}{94}},
  \bibinfo{pages}{070407} (\bibinfo{year}{2005}).

\bibitem[{\citenamefont{Bergli et~al.}(2009)\citenamefont{Bergli, Galperin, and
  Altshuler}}]{bergli2009decoherence}
\bibinfo{author}{\bibfnamefont{J.}~\bibnamefont{Bergli}},
  \bibinfo{author}{\bibfnamefont{Y.~M.} \bibnamefont{Galperin}},
  \bibnamefont{and}
  \bibinfo{author}{\bibfnamefont{B.}~\bibnamefont{Altshuler}},
  \bibinfo{journal}{New J. Phys.} \textbf{\bibinfo{volume}{11}},
  \bibinfo{pages}{025002} (\bibinfo{year}{2009}).

\bibitem[{\citenamefont{Charlebois}()}]{url}
\bibinfo{author}{\bibfnamefont{D.}~\bibnamefont{Charlebois}},
  \emph{\bibinfo{title}{Exact numerical simulation of the ornstein-uhlenbeck
  process}},
  \bibinfo{howpublished}{\url{http://www.mathworks.com/matlabcentral/fileexchange}}.

\bibitem[{\citenamefont{Gillespie}(1996)}]{gillespie1996exact}
\bibinfo{author}{\bibfnamefont{D.~T.} \bibnamefont{Gillespie}},
  \bibinfo{journal}{Phys. Rev. E} \textbf{\bibinfo{volume}{54}},
  \bibinfo{pages}{2084} (\bibinfo{year}{1996}).

\bibitem[{\citenamefont{Ithier et~al.}(2005)\citenamefont{Ithier, Collin,
  Joyez, Messon, Vion, Esteve, Chiarello, Shnirman, Makhlin, Schriefl and Schon}}]{ithier2005decoherence}
\bibinfo{author}{\bibfnamefont{G.}~\bibnamefont{Ithier}},
  \bibinfo{author}{\bibfnamefont{E.}~\bibnamefont{Collin}},
  \bibinfo{author}{\bibfnamefont{P.}~\bibnamefont{Joyez}},
  \bibinfo{author}{\bibfnamefont{P.~J.} \bibnamefont{Messon}},
  \bibinfo{author}{\bibfnamefont{D.}~\bibnamefont{Vion}},
  \bibinfo{author}{\bibfnamefont{D.}~\bibnamefont{Esteve}},
  \bibinfo{author}{\bibfnamefont{F.}~\bibnamefont{Chiarello}},
  \bibinfo{author}{\bibfnamefont{A.}~\bibnamefont{Shnirman}},
  \bibinfo{author}{\bibfnamefont{Y.}~\bibnamefont{Makhlin}},
  \bibinfo{author}{\bibfnamefont{A.}~\bibnamefont{Shnirman}},
  \bibinfo{author}{\bibfnamefont{J.}~\bibnamefont{Schriefl}}, \bibnamefont{and}
  \bibinfo{author}{\bibfnamefont{G.}~\bibnamefont{Schon}},
  \bibinfo{journal}{Phys. Rev. B} \textbf{\bibinfo{volume}{72}},
  \bibinfo{pages}{134519} (\bibinfo{year}{2005}).

\end{thebibliography}
\end{document}